\documentclass[twocolumn]{aa} 
\usepackage{graphicx}
\usepackage[breaklinks,colorlinks,citecolor=blue,linkcolor=blue]{hyperref}
\usepackage{txfonts}
\usepackage{lscape}
\usepackage{epstopdf}
\usepackage{CJK}
\usepackage{float}
\usepackage[absolute]{textpos}
\usepackage{amsmath}
\usepackage{mathrsfs}

\begin{document}
\begin{CJK*}{UTF8}{gbsn}
\title{First Detection of Radio Recombination Lines of Ions Heavier than Helium}

\author{Xunchuan Liu \inst{1} \and
       Tie Liu \inst{1} \and
       Zhiqiang Shen \inst{1} \and
       Paul F. Goldsmith \inst{2} \and
       Neal J. Evans II \inst{3} \and
       Sheng-Li Qin \inst{4} \and
       Qiuyi Luo \inst{1} \and
       Yu Cheng \inst{5} \and
       Sheng-Yuan Liu \inst{6} \and
       Fengyao Zhu \inst{7} \and
       Ken'ichi Tatematsu \inst{8} \and
       Meizhu Liu \inst{4} \and
       Dongting Yang \inst{4} \and
       Chuanshou Li \inst{4} \and
       Li Chen \inst{4} \and
       Juan Li \inst{1} \and
       Xing Lu \inst{1} \and
       Qilao Gu \inst{1} \and
       Rongbing Zhao \inst{1} \and
       Bin Li \inst{1} \and
       Yajun Wu \inst{1} \and
       Weiye Zhong \inst{1} \and
       Zhang Zhao \inst{1} \and
       Jinqing Wang \inst{1} \and
       Qinghui Liu \inst{1} \and
       Bo Xia \inst{1} \and
       Li Fu \inst{1} \and
       Zhen Yan \inst{1} \and
       Chao Zhang \inst{1} \and
       Lingling Wang \inst{1} \and
       Qian Ye \inst{1} \and
       Hongli Liu \inst{4} \and
       Chao Zhang \inst{9} \and
       Fengwei Xu \inst{10} \and
       Dipen Sahu \inst{6} 
          }

\institute{Shanghai Astronomical Observatory, Chinese Academy of Sciences, Shanghai 200030, PR China\\
\email{liuxunchuan@shao.ac.cn; liutie@shao.ac.cn; zshen@shao.ac.cn}
\and
Jet Propulsion Laboratory, California Institute of Technology, 4800 Oak Grove Drive, Pasadena CA 91109, USA
\and
Department of Astronomy, The University of Texas at Austin, 2515 Speedway, Stop C1400, Austin, Texas 78712-1205, USA
\and
Department of Astronomy, Yunnan University, Kunming, 650091, PR China
\and
National Astronomical Observatory of Japan, 2-21-1 Osawa, Mitaka, Tokyo, 181-8588, Japan
\and
Institute of Astronomy and Astrophysics, Academia Sinica, Roosevelt Road, Taipei 10617, Taiwan (R.O.C)
\and
Center for Intelligent Computing Platforms, Zhejiang Laboratory, Hangzhou, 311100, PR China
\and
Nobeyama Radio Observatory, National Astronomical Observatory of Japan, National Institutes of Natural Sciences, 462-2 Nobeyama, Minamimaki, Minamisaku, Nagano 384-1305, Japan
\and
Institute of Astronomy and Astrophysics, School of Mathematics and Physics, Anqing Normal University, Anqing, China
\and
Kavli Institute for Astronomy and Astrophysics, Peking University, 5 Yiheyuan Road, Haidian District, Beijing 100871, PR China
}

\abstract{
We report the first detection of radio recombination lines (RRLs) of ions
heavier than helium. 
In a highly  sensitive multi-band (12--50 GHz) line survey toward Orion KL with the TianMa 65-m Radio Telescope (TMRT), we
successfully detected more than fifteen unblended  $\alpha$ lines of RRLs of singly ionized species (\ion{X}{II}) recombined from \ion{X}{III}. 
The Ka-band (35--50 GHz) spectrum also shows tentative signals of 
$\beta$ lines of ions.
The  detected lines can be successfully crossmatched with the the 
rest frequencies of RRLs of \ion{C}{II} and/or \ion{O}{II}.
This finding greatly expands the connotation of  ion RRLs,
since before this work only two blended lines (105$\alpha$ and 121$\alpha$) of \ion{He}{II} had been reported.
Our detected lines  can be fitted simultaneously under assumption of  local thermodynamic equilibrium (LTE). An abundance of \ion{C}{III} and \ion{O}{III} of 8.8$\times$10$^{-4}$ is obtained, 
avoiding the complexities of optical/infrared observations and the blending of RRLs of atoms.
It is consistent with but approaches the upper bound
of the value (10$^{-4}$--$10^{-3}$)  estimated from optical/infrared observations.
The effects of dielectronic recombination may contribute to  enhancing the level populations even at large $n$.
We expect future observations using radio interferometers 
could break the degeneracy between C and O,
and help to reveal the ionization structure and dynamical evolution
of various ionized regions. 
}
\keywords{ISM: HII regions -- Radio lines: ISM -- Line: identification -- Stars: formation -- ISM: abundances  }
\maketitle

\section{Introduction} \label{secintr}

Radio recombination lines (RRLs) are commonly defined as radio spectral lines resulting
from transitions of high-$n$ levels  of atoms, appearing after the recombination of 
singly inoized ions and electrons \citep{2002ASSL..282.....G}. 
This was not a flaw, since so far most of the  detected RRLs are  from transitions of neutral atoms (e.g., H, He and C).
\ion{He}{II} $121\alpha$ and $105\alpha$ in NGC7027 and
NGC 6302 had been reported  
\citep{1976ApJ...210..108C,1980ASSL...80...75T,
1980ASSL...80...81M,1981A&A....96..278W,1987RMxAA..14..560G,
1990A&A...230..457V}.
However, detection of RRLs of  ions with mass larger than He has never been reported. Searching for RRLs of ions heavier than helium  
towards the Sun has also been unsuccessful
\citep{1972ApJ...171..191B,2022ARep...66..490D}. 

\begin{figure}[!thb]
\centering
\includegraphics[width=0.7\linewidth]{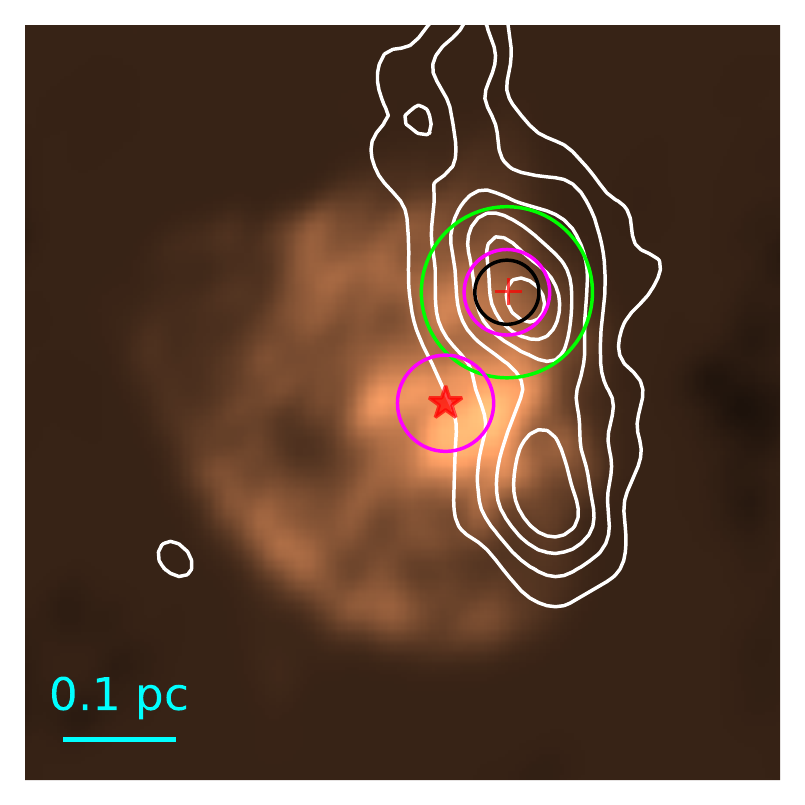}
\caption{Contours of SCUBA 850 $\mu$m dust emission \citep[Orion KL;][]{2008ApJS..175..277D} 
overlaid on the 6 cm VLA continuum image of the M42 \ion{H}{II} region. The red cross is IRc2 of Orion KL (RA(J2000)=05:35:14.55, 
DEC(J2000)=$-$05:22:31.0).
The red pentagram denotes $\theta^1$ Ori C 
(RA(J2000)=5:35:16.47, DEC(J2000)=$-$5:23:22.9), an O6-type star that is 
the dominant ionization source of M42 \citep{2000AJ....120..382O}.
The pink, green, and black circles
represent the beams of TMRT in Ka, Ku and Q bands, respectively (Section \ref{secobs}).
\label{figorionkl}}
\end{figure} 

RRLs have unique advantage in studying ionized gas compared with
optical/infrared observations. 
The complexities of non-LTE population, fluctuations
in temperature and density, and non-negligible extinction 
often face to optical/infrared observations \citep[e.g.,][]{1967ApJ...150..825P,
1991ApJ...366..107F,2001MNRAS.323..343L}. 
These complexities can be easily avoided by RRLs.
For example, the line ratio between RRLs of H and He could provide a direct 
measurement of He abundance, merely weakly dependent on temperature and
density \citep{2002ASSL..282.....G,2009ApJ...690..706A,2011ApJS..194...32A}.
Unfortunately, the study of heavier elements in ionized regions
usually can not take advantage of the benefits of RRLs. 
RRLs of atoms heavier than helium, including C and O which are the most important constituents of CO and interstellar complex organic molecules, have 
rest frequencies close to He RRLs, and
the line blending makes them difficult to be spectrally resolved \citep[e.g.,][]{2019A&A...626A..70S,2021RAA....21..209Z}. 
The RRLs of ions are expected not to be blended with RRLs of neutral atoms,
and thus measuring them would be extremely valuable for constraining the properties of elements heavier than helium in ionized states.
Unlike \ion{He}{II}, ions heavier than helium are usually  multi-electron
systems. Thus, RRLs of ions heavier than helium would also be very  important for studying 
the mechanisms of recombination and level population \citep[e.g.;][]{1981MNRAS.195P..27S,1983A&A...126...75N,2019ApJ...887L...9N}.


So, could we detect  RRLs of ions heavier than helium?
From the multi-band  spectrum of our on-going multi-band (35--50 GHz) TianMa 65-m Radio Telescope (TMRT) line survey
of Orion KL, we successfully
matched firm signals of RRLs of \ion{C}{II} and \ion{O}{II}. 
In this paper, we report the first detection of radio recombination lines (RRLs) of ions heavier than helium in 
Orion KL, and, to our knowledge, also in the interstellar medium.

\begin{table}[!thb]
\caption{Gaussian parameters of ion RRLs of Orion KL. \label{tab_gau}}
\begin{tabular}{ccccc}
\hline
Transition\tablefootmark{(1)}
             &  $f_0$\tablefootmark{(2)}      
&  V$_{\rm LSR}$\tablefootmark{(3)} 
&  $\Delta V$ &  $T_{\rm peak}$\tablefootmark{(4)} \\
             &  (MHz)        & (km s$^{-1}$)& (km s$^{-1}$) &  (mK)         \\   
\hline
91$\alpha$   & 34356.4150  & -10(1)         &   11(3)        &  7(2) \\
92$\alpha$   & 33254.1129  & -10(2)         &   19(3)        &  6(2) \\
95$\alpha$\tablefootmark{(5)}   
             & 30217.5243  & -6.5(1)         &   13(1)        &  11(2)          \\
97$\alpha$   & 28395.8047  & -9(1)       &   15(1)        &  12(2)         \\
98$\alpha$   & 27539.6791  &  -7.5(1)        &   15(1)        &  11(2)         \\
98$\alpha$\tablefootmark{(6)}   & 27539.6791  & -3.5(1)       & 17(1)      &  13(2) \\
99$\alpha$   & 26717.6266  &  -8(1)        &   15(1)        &  11(2)         \\
117$\alpha$  & 16223.6031  &  -7(0.5)    &   12(1)        &  20(2) \\
118$\alpha$\tablefootmark{(7)}
             & 15816.3257  & -6(0.5)     &   13(1)        &  18(2) \\
119$\alpha$  & 15422.5674  & -6(0.5)     &   15(1)        &  21(3) \\
120$\alpha$  & 15041.7719  & -6(1)     &   14(1)        &  16(3) \\
121$\alpha$\tablefootmark{(8)}  & 14673.4103  & -6(1)     &   13(2)        &  18(3) \\
122$\alpha$  & 14316.9791  & -5(0.5)     &   18(1)        &  25(3) \\
123$\alpha$  & 13971.9991  & -5(0.5)     &   15(1)        &  25(4) \\
124$\alpha$  & 13638.0143  & -5(0.5)     &   13(1)        &  24(4) \\
\hline
\end{tabular}\\
{\small
\tablefoottext{1}{Only RRLs of \ion{X}{II} in Ka/Ku band which are unblended or slightly blended are listed.}
\tablefoottext{2}{The rest frequencies of RRLs of \ion{C}{II} (equation \ref{eq_rrlfreq}) are listed.}
\tablefoottext{3}{The velocities will be 3.5 km s$^{-1}$ redder than the values listed here
if the rest frequencies of RRL of \ion{O}{II} are adopted (section \ref{sec_rrl}).
The numbers in brackets in the 3rd to 5th columns 
represent the uncertainties.}  
\tablefoottext{4}{The uncertainty of $T_{\rm peak}$ is the 1-$\sigma$ noise at a velocity resolution of $\sim$1 km s$^{-1}$.}
\tablefoottext{5}{Weakly blended with emission of SO$_2$.}
\tablefoottext{6}{This row is for the observation towards $\theta^1$ Ori C.}
\tablefoottext{7}{A transition of CH$_3$OCH$_3$ has close frequency but can be reasonably ignored (Figure \ref{XIIrrl_kafig}).}
\tablefoottext{8}{partly blended with He160$\kappa$ and H195$\tau$.}
}
\end{table}

\section{Observations}\label{secobs}
Data in this work are mainly from an on-going TMRT line survey toward Orion KL (Figure \ref{figorionkl}),
which aims to  cover the whole frequency range (1--50 GHz) of TMRT, began with
the Q-band survey  \citep[34.8--50 GHz;][]{2022ApJS..263...13L}.
Observations of the Ka-band (26--35 GHz) survey have  been finished  (Liu et al. in prep.).
The Ku-band survey (12--18 GHz) has been partly executed.

In the Ka-band line survey, the mode 2 of the spectral backend (DIBAS) was adopted,
which provides two independent frequency banks of 1.5 GHz, 
with each polarization of each bank having 16384 channels, 
corresponding to a frequency resolution of 91.553 kHz ($\sim$0.92 km s$^{-1}$ at 30 GHz). 
The position switching observation mode was adopted. 
The off-source position is 0.25\degr~away (in azimuth direction) 
from the target. 
We shifted the frequency of the spectrometer banks  to cover 
26--35 GHz. The spectra were then chopped into segments of $\sim$100 MHz in 
frequency bandwidth. We manually fitted and subtracted the baselines to those
segments before splicing them to obtain the final spectrum of Orion KL.
For comparison, we also observed  towards
$\theta^1$ Ori C with one frequency setup 
in Ka band, covering the \ion{X}{II} 98$\alpha$ transition (Table \ref{tab_gau}).

The Ku-band observations also serve as part of the TMRT line survey. 
The frequency ranges covering the expected $\alpha$ lines of ions were
preferentially observed.
Mode 3 of the DIBAS was adopted, 
which provides two independent frequency banks of 500 MHz
with  a frequency resolution of 30.517 kHz ($\sim$0.61 km s$^{-1}$ at 15 GHz). 

For calibration, the signal from a noise diode is periodically injected.
Under typical weather conditions at the TMRT in winter with an air pressure of 
1000 mbar and a water vapor density of 8 g m$^{-3}$,
the zenith atmospheric opacities are 0.03 and 0.1 in the Ku and Ka band, respectively 
\citep{2017AcASn..58...37W}.
Calibration uncertainties are estimated to be
less than 20\% \citep{2017AcASn..58...37W,2022ApJS..263...13L}.


\begin{figure*}[!htp]
\centering
\includegraphics[width=0.82\linewidth]{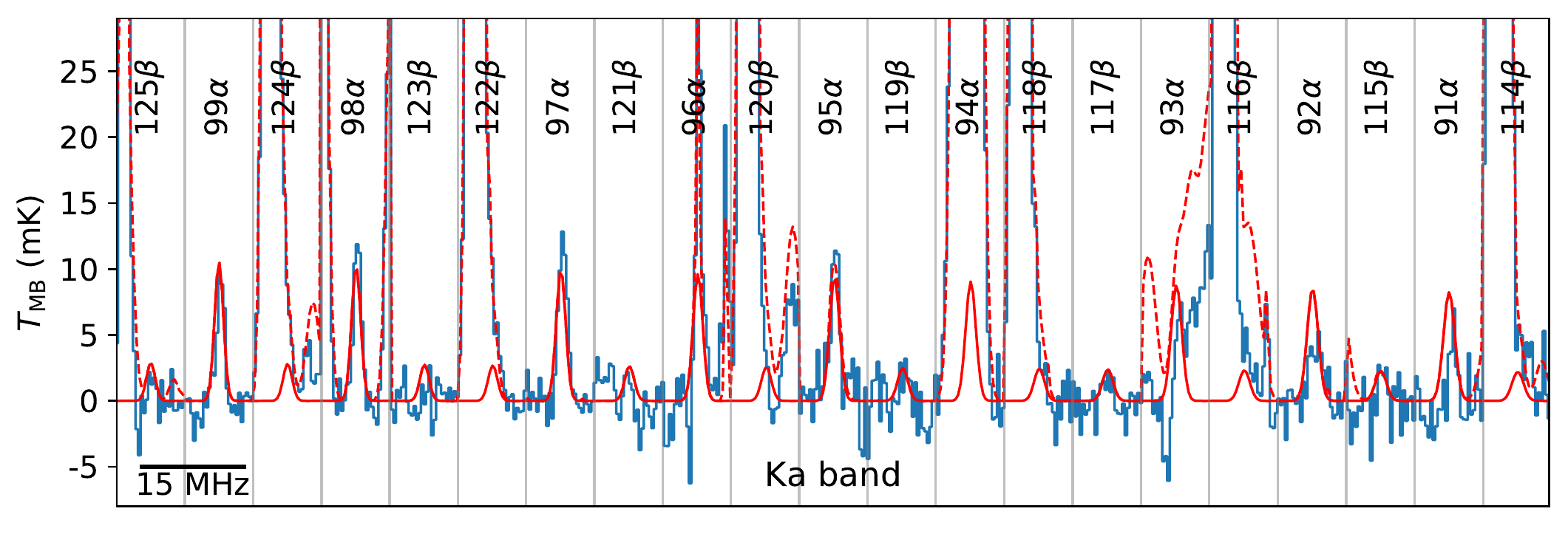}
\includegraphics[width=0.82\linewidth]{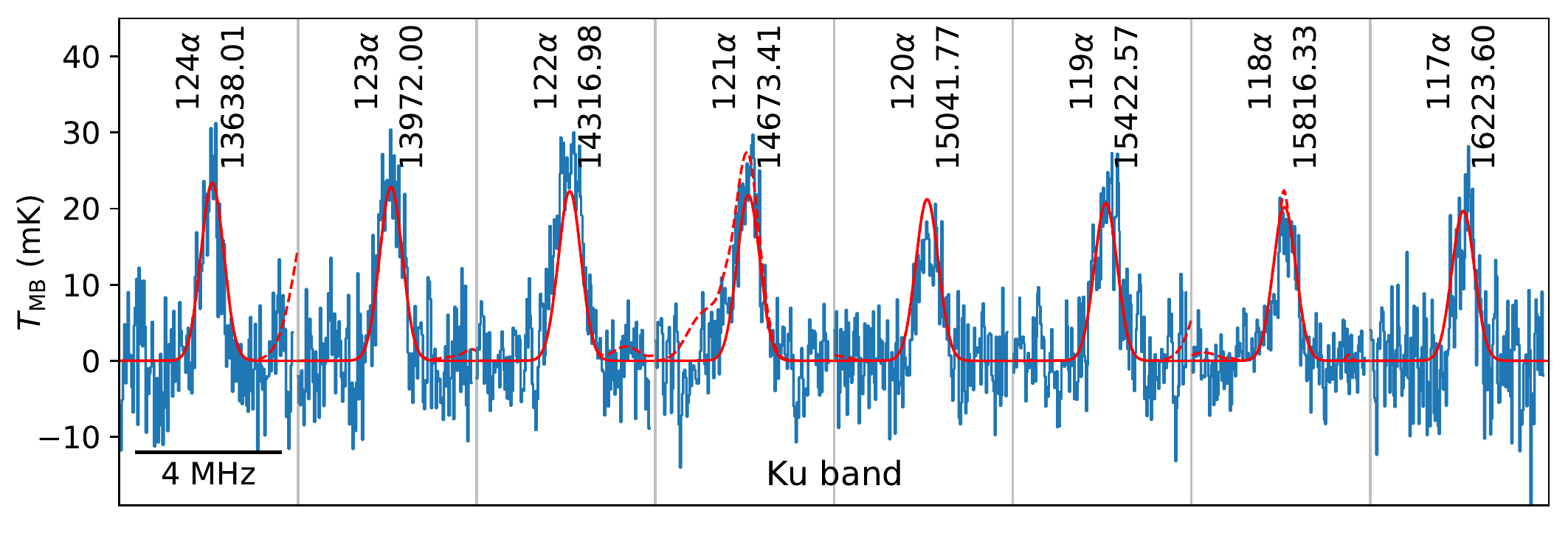}
\includegraphics[width=0.82\linewidth]{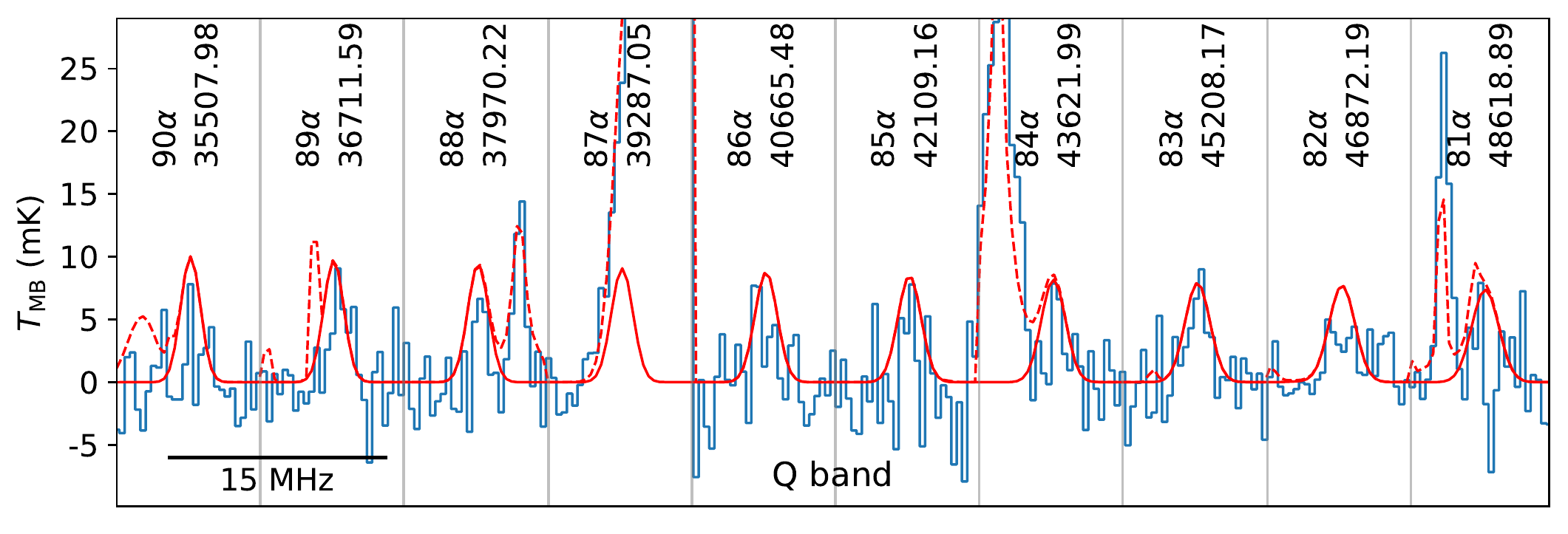}
\caption{The spliced spectrum (blue line) of Orion KL around the rest frequencies of
RRLs of ions (\ion{C}{II} or \ion{O}{II}) in Ka band (upper panel), Ku band (middle panel) and Q band (lower panel). 
The Ka/Q spectra have been smoothed to have a spectral resolution ($\Delta \nu$) of 366.212 kHz.
The Ku spectrum is unsmoothed with a $\Delta \nu$ of 30.517 kHz.
Segments  separated by gray vertical lines cover different 
frequency ranges.
The transition labels and the rest frequencies of the corresponding
\ion{C}{II} transitions (in unit of MHz) are shown on the top axis.
The solid red line represents the model fitting of \ion{X}{II} RRLs (Table \ref{tab_rrldp}).
The dashed line includes the contributions of all the RRLs as well as all transitions of molecules 
identified in the Q-band survey \citep{2022ApJS..263...13L} and Ka-band survey.
\label{XIIrrl_kafig}
}
\end{figure*}


\section{Identification of ion RRLs} \label{sec_rrl} \label{sectid}
For a hydrogenic emitter with a total mass of $M$ and a total charge of $Z-1$ 
(where $Z$ is the atomic charge of the species that has just recombined), the rest frequency of an RRL can be 
expressed as \citep{2002ASSL..282.....G}
\begin{equation} \label{eq_rrlfreq}
\nu^{\rm RRL}_{\rm rest}(n+\Delta n,n) = R cZ^2
\left( \frac{1}{n^2}-\frac{1}{(n+\Delta n)^2} \right).
\end{equation} 
with Rydberg constant $R$ as \citep{1996ApJS..107..747T}
\begin{equation} 
R = R_\infty \frac{M-Zm_e}{M-(Z-1)m_e}.
\end{equation}
Here, $c$ is the  speed of light, $R_\infty=
109737.31568$ cm$^{-1}$ \citep{2021RvMP...93b5010T}, $m_e$
is the mass of electron, and $M$ is the mass of the corresponding neutral atom.
The Rydberg constants (in cm$^{-1}$) for H, He, C, \ion{He}{II} (He$^+$), \ion{C}{II} (C$^+$), and \ion{O}{II} (O$^+$) are
109677.58, 109722.28, 109732.30,  109722.27,  109732.30, and 109733.55 respectively. 
The $n$ is large for RRLs, and it is thus valid to
treat atoms and ions  as hydrogenic emitters \citep{1972ApJ...171..191B}
even through they may be not hydrogenic when $n$ is small  \citep{2022MNRAS.513.1198D}.
The factor of $Z^2$ could separate these lines from those of the neutral species, avoiding the blending issues described above.


\begin{figure*}[!t]
\includegraphics[width=0.49\linewidth]{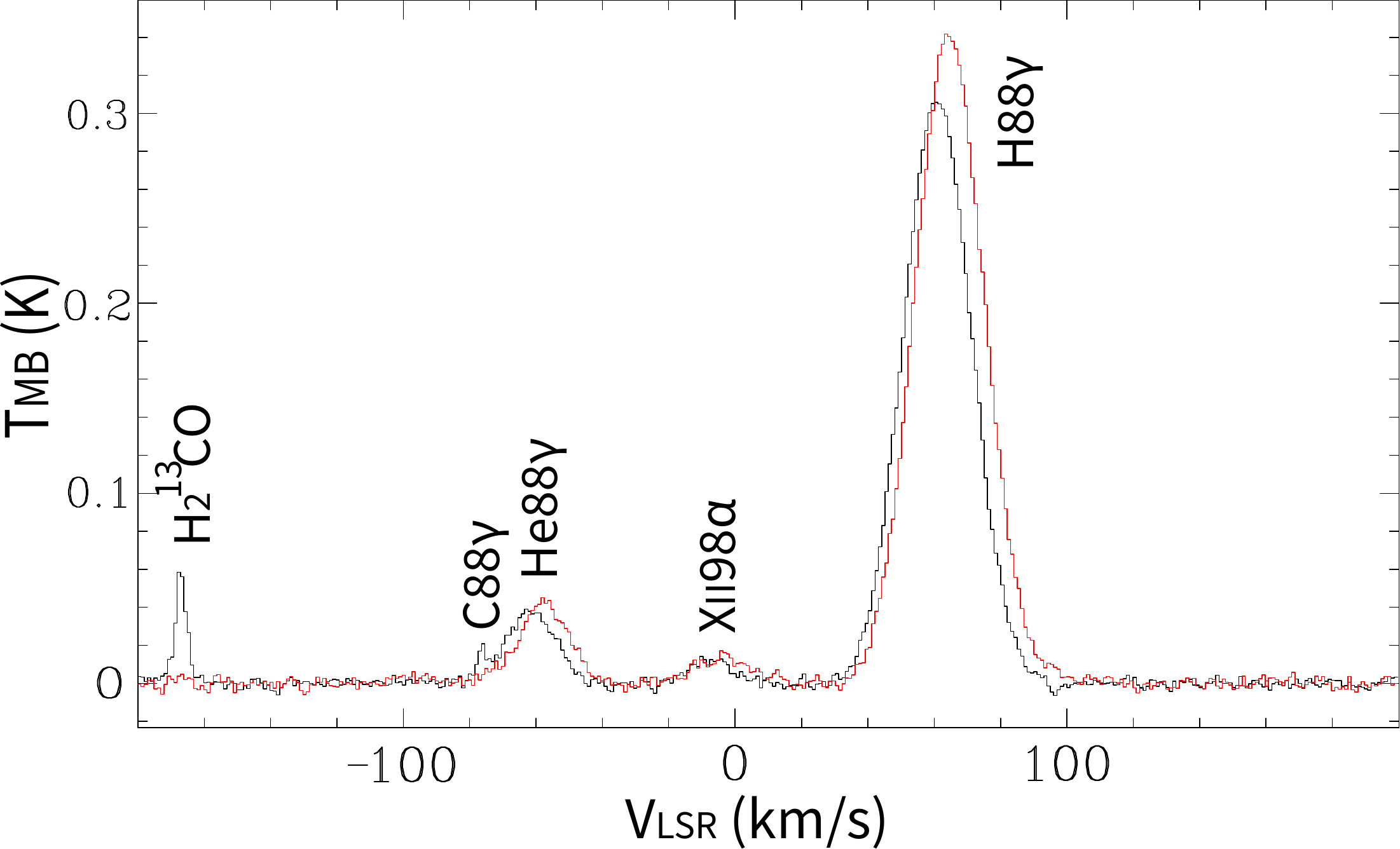}
\includegraphics[width=0.49\linewidth]{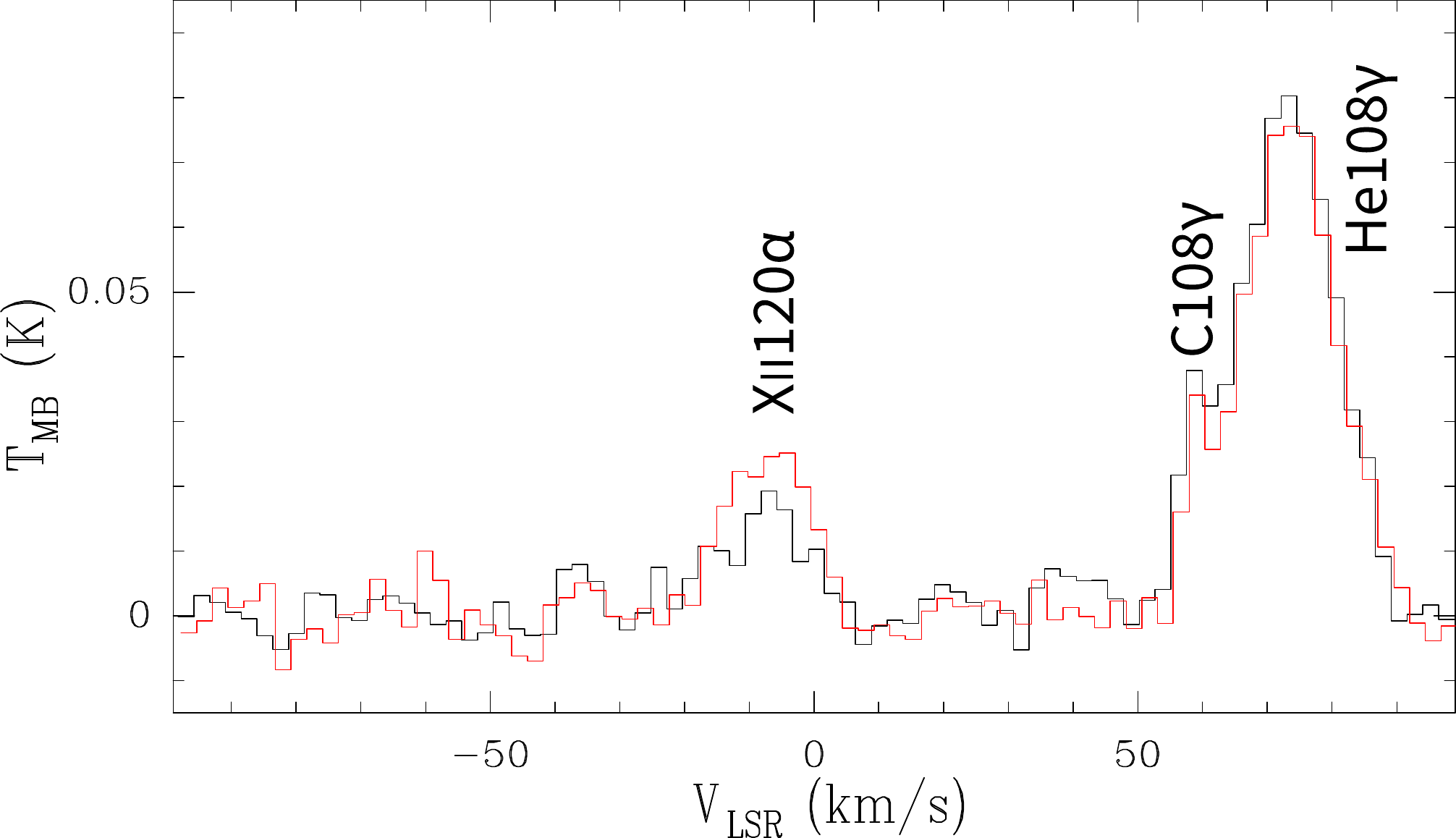}
\caption{Left: Comparison between the the spectra towards Orion KL (black) and towards $\theta^1$ Ori C (red). 
Right: Comparison between the spectra towards Orion KL observed on different days. \label{fig_comp}}
\end{figure*}

\subsection{Line matching}
During modeling the Ka-band  spectrum of Orion KL following the procedure of \citet{2022ApJS..263...13L},
we found six clean and broad ($\Delta V>$10 km s$^{-1}$) line features that cannot be 
assigned to any RRLs of atoms (H, He and C) or molecular lines. 
Instead, we successfully matched them
with  the frequencies of  RRLs of \ion{C}{II} calculated by Equation \ref{eq_rrlfreq}.
Since the equivalent velocity offset between the same RRL transitions of oxygen and carbon is only $\sim$3.5 km s$^{-1}$, 
much smaller than the typical line widths of RRLs, 
we cannot distinguish them at this stage and the emitter of detected ion RRLs is thus denoted as X.   
Further, we successfully detected eight $\alpha$ lines of \ion{X}{II} in follow-up Ku-band  observations.  
The spectrum of Q-band survey has lower line sensitivity (higher noise) compared to the Ka-band one. 
Several $\alpha$ lines ($n=$83, 84, 88, 89) are marginally detected in Q band.
The \ion{X}{II}98$\alpha$ are also firmly detected towards $\theta^1$ Ori C (left panel of Figure \ref{fig_comp}).
The results of the Gaussian fitting of the unblended $\alpha$ lines of \ion{X}{II} are listed in 
Table \ref{tab_gau}. 

For the $\beta$ lines of \ion{X}{II},  those with an even number of $n$  have a  rest frequency identical with that of
\ion{X}{I}(n/2)$\alpha$, and will be highly blended with He(n/2)$\alpha$.
The $\beta$ lines of \ion{X}{II} with an odd number of $n$ are 
unblended. The data do not contradict the model (Section \ref{sec_modfit}), but the lines can only be tentatively
detected under the sensitivity of this survey (Figure \ref{XIIrrl_kafig}).

The ion RRLs of different transitions are usually observed on different days,
expanding a period of several months. The Doppler shifts
introduced by earth revolution varied by tens of km s$^{-1}$,
much larger than the velocity shifts of detected lines.
For example, we observed the \ion{X}{II}120$\alpha$ towards Orion KL
on two different days 
with a difference of Doppler shifts of 15 km s$^{-1}$,
and the velocities of the detected lines (in $V_{\rm LSR}$)
remain unchanged (right panel of Figure \ref{fig_comp}).
The ion RRLs should originate from interstellar space.

\subsection{Model fitting} \label{sec_modfit}
The  RRLs, including RRLs of ions, could be modeled following Equations 4--8 of of \citet{2022ApJS..263...13L}.
The peak value of optical depth can be derived as \citep{2002ASSL..282.....G}
\begin{equation}\label{kappa_RRL}
\tau_{n_1,n_2} = 3.867\times 10^{-12} \frac{b_{n_2}}{\Delta \nu}  \frac{\Delta n}{n_1}Z^2 f_{n_1,n_2}
\left(1-\frac{3\Delta n}{2n_1}\right) \frac{EM}{T_e^{5/2}}.
\end{equation}
Here,
$EM$ is the emission measure  ($\int n_in_e dl$) in unit of 
cm$^{-5}$, $n_{i}$ is the density of $X^{\rm Z+}$, 
$n_{e}$ is the  density of free electrons, 
$n_1$ ($n_2$) is the quantum number of the lower (upper) level,
$b_{n_2}$ is the upper-level departure coefficient, 
$\Delta \nu$ is the line width in unit of Hz,
and $T_e$ is the excitation temperature.
The oscillator strength $f_{n_1,n_2}$
is independent of Z \citep{1959SvA.....3..813K,1968ApJS...17..445G}.

For RRLs from \ion{H}{II} regions (RRLs of
H, He and \ion{X}{II}), the electron temperature ($T_{\rm e}$) is 
adopted as 8000 K \citep{1997A&A...327.1177W,2006MNRAS.370..597L}.
For RRLs of \ion{X}{II}, 
the rest frequencies of \ion{C}{II} RRLs are adopted (Equation \ref{eq_rrlfreq}).
We assume that the high-$n$ levels are in LTE  ($b_n=1$).
This assumption is valid for H RRLs of Orion KL \citep{2022A&A...665A..94Z,2022ApJS..263...13L}.
Since the $n$ tend to be larger for \ion{X}{II} RRLs, 
this assumption should also be valid for \ion{X}{II}.
For comparison, we also fit the carbon RRL from PDR assuming a $T_{\rm e}$ of 300 K
\citep{2022A&A...658A..98P}.
The Ku/Ka/Q-band line features can be well fitted simultaneously,
suggesting that the RRLs of both atoms and ions are indeed in LTE.
The fitted results are shown in Table \ref{tab_rrldp}
and Figure \ref{XIIrrl_kafig}.

The  fitted velocity of \ion{X}{II} RRL is -5.5 km s$^{-1}$, assuming X to be carbon.
If X is assumed to be oxygen, the fitted velocity will be -2 km s$^{-1}$.
These two values are both close to the velocity of -4 km s$^{-1}$
derived from RRLs of H and He (Table \ref{XIIrrl_kafig}), and the velocity
differences are much smaller than the line widths ($\sim$15 km s$^{-1}$).
Since the Ku-band lines tend to be systematically $\sim$2 km s$^{-1}$ redder than those in 
Ka band, the uncertainty of fitted $V_{\rm LSR}$ is estimated to be 2 km s$^{-1}$. This may be leaded by the larger beam at Ku band which 
thus covers more \ion{C}{III} than \ion{O}{III} (Section \ref{secdistribution}).

We also fitted the parameters of RRLs  of $\theta^1$ Ori C (Table \ref{tab_rrldp}).
The intensities of RRLs at Orion KL and $\theta^1$ Ori C are similar.
The velocities of RRLs of H and He at $\theta^1$ Ori C are $\sim4$ km s$^{-1}$ redder than  at Orion KL.
The \ion{X}{II} RRLs at $\theta^1$ Ori C are $\sim 2$ km s$^{-1}$ redder and broader than at
Orion KL. 

\begin{table}[!thb]
\caption{The model parameters of RRLs \label{tab_rrldp}}
\begin{tabular}{ccccc}
\hline
 & region & $EM$ & $V_{\rm LSR}$\tablefootmark{(1)} & $\Delta V$ \\
 &        & (cm$^{-6}$ pc) & (km s$^{-1}$) & (km s$^{-1}$)\\
\hline
H  & M42 (Orion KL)   & 1.46$\times$10$^{6}$ &   -4            &    25        \\
He & M42 (Orion KL)   & 1.46$\times$10$^{5}$ &   -4            &    17.5      \\
\ion{C}{II}(\ion{O}{II}) & M42 (Orion KL)   & 1.29$\times$10$^{3}$ & -5.5(-2)            &    15  \\
C\tablefootmark{(2)} & PDR   & 1.80$\times$10$^{2}$ &   8            &    5      \\
H  & M42 ($\theta^1$ Ori C)   & 1.7$\times$10$^{6}$ &   0            &    24        \\
He  & M42 ($\theta^1$ Ori C)   & 1.7$\times$10$^{5}$ &   0            &    16.5        \\
\ion{C}{II}(\ion{O}{II}) & M42 ($\theta^1$ Ori C)   & 1.48$\times$10$^{3}$ & -3.5(0)            &    17  \\
\hline
\end{tabular}\\
{\small
\tablefoottext{1}{The values in the brackets are for \ion{O}{II}.}
\tablefoottext{2}{A $T_e$ of 300 K is adopted for carbon RRLs from PDRs.} 
}
\end{table}

\subsection{Ion abundance and distribution}  \label{secionab}\label{secdistribution}
The extinction by absorption of dust and self-absorption can usually be
neglected for RRLs. Thus, the abundance ratio of two ions can be simply derived from the intensity ratio of their RRLs.
The fitted n(X$^{2+}$)/n(H$^+$), derived through $EM($\ion{X}{II}$)/EM({\rm H})$, is 8.8$\times$10$^{-4}$ at both Orion KL and $\theta^1$ Ori C.
It is close to the
solar abundances of C and O 
\citep[8$\times$10$^{-4}$;][]{2021A&A...656A.113A}
and approaches the upper bound
of the value of Orion Nebula  estimated from optical/infrared observations  \citep[10$^{-4}$--10$^{-3}$;][]{1986ApJ...311..895S,1993ApJ...414..626P,1998MNRAS.295..401E}. 
This may hint that the effects of dielectronic recombination could enhance the level populations of 
multi-electron ions at $n$ even as high as $\sim$100
\citep{1981MNRAS.195P..27S,2019ApJ...887L...9N}.


Helium can be excluded as the emitter of the detected RRLs, since it would give too large a value of 
$V_{LSR}$ of -33 km s$^{-1}$.
In addition, the high ionization energy of \ion{He}{II}, 54.418 eV, would lead to a
very small ratio of R(\ion{He}{III})/R(\ion{He}{II}) of $\sim 0.02$ 
for an O6 type star \citep{2008MNRAS.389.1009S},
and the \ion{He}{III} region would be highly beam diluted in our observations (Figure \ref{figorionkl}).
The \ion{X}{III} abundance is much higher than the total abundance of nitrogen.
Thus,  the contribution of \ion{N}{II} RRLs  can also be ignored. 

The ionization energy of \ion{C}{II}, 24.383 eV, is close to the value of
\ion{He}{I}, 24.587 eV. Thus, the distribution of \ion{C}{III} is expected to be as extended 
as \ion{He}{II}, as revealed by optical observations
\citep{1991PASP..103..830W,1998MNRAS.295..401E}. However, the contribution of oxygen can not be  excluded.
The fitted velocities of Ka-band lines at Orion KL and
Ku-band lines at $\theta^1$ Ori C are closer to the
values of H/He RRLs if they are assigned to oxygen ions.  
It is also consistent 
with the value derived from the \ion{O}{III} optical lines  \citep[V$_{LSR}\sim 0\pm 3$ km s$^{-1}$;][]{2019ApJ...881..130A}. 
On the contrast, the fitted velocities of Ku-band  lines at Orion KL 
are closer to the
values of H/He RRLs if carbon is adopted. 
This is consistent with the scenario that the inner part of the \ion{X}{III} region of
M42 is dominated by \ion{O}{III}, while \ion{C}{III}
dominates  the outer part, since the ionization energy of O$^{n+}$ is 
larger than of C$^{n+}$.

\section{Discussion} \label{secfuture}\label{secdis}
As mentioned in Section \ref{secintr}, previously only two lines of \ion{He}{II} 
have been reported.
The rest frequency of \ion{He}{II} $121\alpha$ 
(14672.069 MHz) is coincident with that of He160$\kappa$
(14672.078 MHz), and
close to that of H$195\tau$ (14672.993 MHz).
The \ion{He}{II}$105\alpha$ (22411.329 MHz) is 
blended with He$129\theta$ (22411.276 MHz).
In ionized regions with very hard ionizing spectrum, those 
\ion{He}{II} lines could overwhelm the blending high-order He lines.
However, the intensity of those \ion{He}{II} lines may be weaker than 
the He lines  for  ionized regions such as the Orion Nebula (Figure
\ref{XIIrrl_kafig}).
Further observations and modeling are probably needed to examine the 
blending issue of the \ion{He}{II} RRLs lines.
This work detected tens of lines of RRLs of carbon and oxygen, many of them are unblended, which will be very helpful for further studies of ion RRLs in both observation and theory.

Future high spectral and spatial resolution mapping observations of RRLs of ions in different ionization states with the state-of-the-art interferometers such as ALMA (band 1), SKA or ngVLA can reveal the ionization structure
and break the degeneracy between C and O. This would
 measure the distributions and abundances of these ions, 
 and consequently elements heavier than helium directly and separately. 
Such a technique would be very valuable to study the abundances of
carbon and oxygen in the inner Galaxy, where optical observations are very difficult. Studies with optical lines are limited to $R_{gal} > 5$ kpc for oxygen and $R_{gal} > 6$ kpc for carbon
\citep{2022MNRAS.510.4436M}.
The abundance of C and O affects the conversion of CO luminosity 
into molecular gas mass 
\citep{2020ApJ...903..142G, 2022ApJ...931...28H}.
Higher abundances in the inner Galaxy, along with higher temperatures, will lower the molecular gas masses and consequently the predicted star formation rates in the inner Galaxy,  
alleviating the long-standing discrepancy between predicted and observed star formation rates in the inner Galaxy
\citep{2022ApJ...929L..18E}.
In addition, RRLs of ions tend to have smaller thermal line widths, and the RRLs of ions may serve as a good tracer of gas motions inside \ion{H}{II} regions. This will be helpful to constrain the structures and dynamical evolution models for highly embedded \ion{H}{II} regions.
Observations of these RRLs of ions in deeply embedded ultra-compact \ion{H}{II} regions, where the electron density is much higher than in the Orion Nebula,
could help to constrain 
the theory of collisional broadening of Rydberg transitions 
in ions \citep[e.g.,][]{2021A&A...651A..35O}.

\section{Summary} \label{secsum}
We successfully detected and identified RRLs of ions heavier than helium in the insterstellar meidum for the first time, during the  conduction of the on-going TMRT multi-band (12--50 GHz) line survey towards Orion KL. 
More than fifteen unblended  $\alpha$ lines of RRLs of \ion{C}{II} and/or \ion{O}{II} 
are detected. 
The sensitive Ka-band spectrum even shows tentative signals of $\beta$ lines.
All the detected lines can be well fitted simultaneously under the assumption of 
LTE, yielding an abundance of  \ion{C}{III} and/or \ion{O}{III} of 8.8$\times$10$^{-4}$.
The RRL of ions are extremely useful, serving as a direct and model-independent  measurement  of the abundances of elements heavier than helium. 

The ion RRLs at Orion KL and $\theta^1$ Ori C have similar intensities but slightly shifted
velocities of $\sim$2 km s$^{-1}$.
We  prefer the interpretation that the detected lines towards Orion KL are blended  RRLs of \ion{C}{II} and \ion{O}{II},
while those towards $\theta^1$ Ori C may be dominated by \ion{O}{II} RRLs. 
We expect that future  ion RRL observations using interferometers   with better resolution and sensitivities  can 
 reveal in detail the ionization structures of  ionized regions in widespread environments.

\begin{acknowledgements} \small
We wish to thank the staff of the TMRT 65 m for their
help during the observations. This work has been supported by the National Key R\&D Program of China (No. 2022YFA1603100).
X.L. acknowledges the supports by NSFC No. 12203086 and No. 12033005 and CPSF No. 2022M723278. T.L. acknowledges the supports by National Natural Science Foundation of China (NSFC) through grants No.12073061 and No.12122307, the international partnership program of Chinese Academy of Sciences through grant No.114231KYSB20200009, Shanghai Pujiang Program 20PJ1415500 and the science research grants from the China Manned Space Project with no. CMS-CSST-2021-B06.
This research was carried out in part at the Jet Propulsion Laboratory, which is operated by the California Institute of Technology under a contract with the National Aeronautics and Space Administration
(80NM0018D0004).
We show warm thanks to the anonymous referee for providing many deep-insight comments for improving the paper.
\end{acknowledgements}

\bibliography{ms}

\begin{thebibliography}{44}
\expandafter\ifx\csname natexlab\endcsname\relax\def\natexlab#1{#1}\fi

\bibitem[{{Abel} {et~al.}(2019){Abel}, {Ferland}, \&
  {O'Dell}}]{2019ApJ...881..130A}
{Abel}, N.~P., {Ferland}, G.~J., \& {O'Dell}, C.~R. 2019, \apj, 881, 130

\bibitem[{{Amarsi} {et~al.}(2021){Amarsi}, {Grevesse}, {Asplund}, \&
  {Collet}}]{2021A&A...656A.113A}
{Amarsi}, A.~M., {Grevesse}, N., {Asplund}, M., \& {Collet}, R. 2021, \aap,
  656, A113

\bibitem[{{Anderson} \& {Bania}(2009)}]{2009ApJ...690..706A}
{Anderson}, L.~D. \& {Bania}, T.~M. 2009, \apj, 690, 706

\bibitem[{{Anderson} {et~al.}(2011){Anderson}, {Bania}, {Balser}, \&
  {Rood}}]{2011ApJS..194...32A}
{Anderson}, L.~D., {Bania}, T.~M., {Balser}, D.~S., \& {Rood}, R.~T. 2011,
  \apjs, 194, 32

\bibitem[{{Berger} \& {Simon}(1972)}]{1972ApJ...171..191B}
{Berger}, P.~S. \& {Simon}, M. 1972, \apj, 171, 191

\bibitem[{{Chaisson} \& {Malkan}(1976)}]{1976ApJ...210..108C}
{Chaisson}, E.~J. \& {Malkan}, M.~A. 1976, \apj, 210, 108

\bibitem[{{Del Zanna} \& {Storey}(2022)}]{2022MNRAS.513.1198D}
{Del Zanna}, G. \& {Storey}, P.~J. 2022, \mnras, 513, 1198

\bibitem[{{Di Francesco} {et~al.}(2008){Di Francesco}, {Johnstone}, {Kirk},
  {MacKenzie}, \& {Ledwosinska}}]{2008ApJS..175..277D}
{Di Francesco}, J., {Johnstone}, D., {Kirk}, H., {MacKenzie}, T., \&
  {Ledwosinska}, E. 2008, \apjs, 175, 277

\bibitem[{{Dravskikh} \& {Dravskikh}(2022)}]{2022ARep...66..490D}
{Dravskikh}, A.~F. \& {Dravskikh}, Y.~A. 2022, Astronomy Reports, 66, 490

\bibitem[{{Esteban} {et~al.}(1998){Esteban}, {Peimbert}, {Torres-Peimbert}, \&
  {Escalante}}]{1998MNRAS.295..401E}
{Esteban}, C., {Peimbert}, M., {Torres-Peimbert}, S., \& {Escalante}, V. 1998,
  \mnras, 295, 401

\bibitem[{{Evans} {et~al.}(2022){Evans}, {Kim}, \&
  {Ostriker}}]{2022ApJ...929L..18E}
{Evans}, N.~J., {Kim}, J.-G., \& {Ostriker}, E.~C. 2022, \apjl, 929, L18

\bibitem[{{Fich} \& {Silkey}(1991)}]{1991ApJ...366..107F}
{Fich}, M. \& {Silkey}, M. 1991, \apj, 366, 107

\bibitem[{{Goldwire}(1968)}]{1968ApJS...17..445G}
{Goldwire}, Henry~C., J. 1968, \apjs, 17, 445

\bibitem[{{Gomez} {et~al.}(1987){Gomez}, {Rodriguez}, \&
  {Garcia-Barreto}}]{1987RMxAA..14..560G}
{Gomez}, Y., {Rodriguez}, L.~F., \& {Garcia-Barreto}, J.~A. 1987, \rmxaa, 14,
  560

\bibitem[{{Gong} {et~al.}(2020){Gong}, {Ostriker}, {Kim}, \&
  {Kim}}]{2020ApJ...903..142G}
{Gong}, M., {Ostriker}, E.~C., {Kim}, C.-G., \& {Kim}, J.-G. 2020, \apj, 903,
  142

\bibitem[{{Gordon} \& {Sorochenko}(2002)}]{2002ASSL..282.....G}
{Gordon}, M.~A. \& {Sorochenko}, R.~L. 2002, {Radio Recombination Lines. Their
  Physics and Astronomical Applications}, Vol. 282 (Berlin: Springer)

\bibitem[{{Hu} {et~al.}(2022){Hu}, {Schruba}, {Sternberg}, \& {van
  Dishoeck}}]{2022ApJ...931...28H}
{Hu}, C.-Y., {Schruba}, A., {Sternberg}, A., \& {van Dishoeck}, E.~F. 2022,
  \apj, 931, 28

\bibitem[{{Kardashev}(1959)}]{1959SvA.....3..813K}
{Kardashev}, N.~S. 1959, \sovast, 3, 813

\bibitem[{{Lerate} {et~al.}(2006){Lerate}, {Barlow}, {Swinyard}, {Goicoechea},
  {Cernicharo}, {Grundy}, {Lim}, {Polehampton}, {Baluteau}, {Viti}, \&
  {Yates}}]{2006MNRAS.370..597L}
{Lerate}, M.~R., {Barlow}, M.~J., {Swinyard}, B.~M., {et~al.} 2006, \mnras,
  370, 597

\bibitem[{{Liu} {et~al.}(2022){Liu}, {Liu}, {Shen}, {Qin}, {Luo}, {Cheng},
  {Gu}, {Zhang}, {Zhu}, {Liu}, {Lu}, {Zhao}, {Zhong}, {Wu}, {Li}, {Zhao},
  {Wang}, {Liu}, {Xia}, {Li}, {Fu}, {Yan}, {Zhang}, {Wang}, {Ye}, {Tatematsu},
  {Liu}, {Shang}, {Xu}, {Lee}, {Zhang}, \& {Dutta}}]{2022ApJS..263...13L}
{Liu}, X., {Liu}, T., {Shen}, Z., {et~al.} 2022, \apjs, 263, 13

\bibitem[{{Liu} {et~al.}(2001){Liu}, {Barlow}, {Cohen}, {Danziger}, {Luo},
  {Baluteau}, {Cox}, {Emery}, {Lim}, \& {P{\'e}quignot}}]{2001MNRAS.323..343L}
{Liu}, X.~W., {Barlow}, M.~J., {Cohen}, M., {et~al.} 2001, \mnras, 323, 343

\bibitem[{{M{\'e}ndez-Delgado} {et~al.}(2022){M{\'e}ndez-Delgado}, {Amayo},
  {Arellano-C{\'o}rdova}, {Esteban}, {Garc{\'\i}a-Rojas}, {Carigi}, \&
  {Delgado-Inglada}}]{2022MNRAS.510.4436M}
{M{\'e}ndez-Delgado}, J.~E., {Amayo}, A., {Arellano-C{\'o}rdova}, K.~Z.,
  {et~al.} 2022, \mnras, 510, 4436

\bibitem[{{Mezger}(1980)}]{1980ASSL...80...81M}
{Mezger}, P.~G. 1980, in Astrophysics and Space Science Library, Vol.~80, Radio
  Recombination Lines, ed. P.~A. {Shaver}, 81--97

\bibitem[{{Nemer} {et~al.}(2019){Nemer}, {Sterling}, {Raymond}, {Dupree},
  {Garc{\'\i}a-Rojas}, {Wang}, {Pindzola}, {Ballance}, \&
  {Loch}}]{2019ApJ...887L...9N}
{Nemer}, A., {Sterling}, N.~C., {Raymond}, J., {et~al.} 2019, \apjl, 887, L9

\bibitem[{{Nussbaumer} \& {Storey}(1983)}]{1983A&A...126...75N}
{Nussbaumer}, H. \& {Storey}, P.~J. 1983, \aap, 126, 75

\bibitem[{{O'Dell} \& {Yusef-Zadeh}(2000)}]{2000AJ....120..382O}
{O'Dell}, C.~R. \& {Yusef-Zadeh}, F. 2000, \aj, 120, 382

\bibitem[{{Olofsson} {et~al.}(2021){Olofsson}, {Black}, {Khouri}, {Vlemmings},
  {Humphreys}, {Lindqvist}, {Maercker}, {Nyman}, {Ramstedt}, \&
  {Tafoya}}]{2021A&A...651A..35O}
{Olofsson}, H., {Black}, J.~H., {Khouri}, T., {et~al.} 2021, \aap, 651, A35

\bibitem[{{Pabst} {et~al.}(2022){Pabst}, {Goicoechea}, {Hacar}, {Teyssier},
  {Bern{\'e}}, {Wolfire}, {Higgins}, {Chambers}, {Kabanovic}, {G{\"u}sten},
  {Stutzki}, {Kramer}, \& {Tielens}}]{2022A&A...658A..98P}
{Pabst}, C.~H.~M., {Goicoechea}, J.~R., {Hacar}, A., {et~al.} 2022, \aap, 658,
  A98

\bibitem[{{Peimbert}(1967)}]{1967ApJ...150..825P}
{Peimbert}, M. 1967, \apj, 150, 825

\bibitem[{{Peimbert} {et~al.}(1993){Peimbert}, {Storey}, \&
  {Torres-Peimbert}}]{1993ApJ...414..626P}
{Peimbert}, M., {Storey}, P.~J., \& {Torres-Peimbert}, S. 1993, \apj, 414, 626

\bibitem[{{Salas} {et~al.}(2019){Salas}, {Oonk}, {Emig}, {Pabst}, {Toribio},
  {R{\"o}ttgering}, \& {Tielens}}]{2019A&A...626A..70S}
{Salas}, P., {Oonk}, J.~B.~R., {Emig}, K.~L., {et~al.} 2019, \aap, 626, A70

\bibitem[{{Sim{\'o}n-D{\'\i}az} \& {Stasi{\'n}ska}(2008)}]{2008MNRAS.389.1009S}
{Sim{\'o}n-D{\'\i}az}, S. \& {Stasi{\'n}ska}, G. 2008, \mnras, 389, 1009

\bibitem[{{Simpson} {et~al.}(1986){Simpson}, {Rubin}, {Erickson}, \&
  {Haas}}]{1986ApJ...311..895S}
{Simpson}, J.~P., {Rubin}, R.~H., {Erickson}, E.~F., \& {Haas}, M.~R. 1986,
  \apj, 311, 895

\bibitem[{{Storey}(1981)}]{1981MNRAS.195P..27S}
{Storey}, P.~J. 1981, \mnras, 195, 27P

\bibitem[{{Terzian}(1980)}]{1980ASSL...80...75T}
{Terzian}, Y. 1980, in Astrophysics and Space Science Library, Vol.~80, Radio
  Recombination Lines, ed. P.~A. {Shaver}, 75--80

\bibitem[{{Tiesinga} {et~al.}(2021){Tiesinga}, {Mohr}, {Newell}, \&
  {Taylor}}]{2021RvMP...93b5010T}
{Tiesinga}, E., {Mohr}, P.~J., {Newell}, D.~B., \& {Taylor}, B.~N. 2021,
  Reviews of Modern Physics, 93, 025010

\bibitem[{{Towle} {et~al.}(1996){Towle}, {Feldman}, \&
  {Watson}}]{1996ApJS..107..747T}
{Towle}, J.~P., {Feldman}, P.~A., \& {Watson}, J. K.~G. 1996, \apjs, 107, 747

\bibitem[{{Vallee} {et~al.}(1990){Vallee}, {Guilloteau}, {Forveille}, \&
  {Omont}}]{1990A&A...230..457V}
{Vallee}, J.~P., {Guilloteau}, S., {Forveille}, T., \& {Omont}, A. 1990, \aap,
  230, 457

\bibitem[{{Walmsley} {et~al.}(1981){Walmsley}, {Churchwell}, \&
  {Terzian}}]{1981A&A....96..278W}
{Walmsley}, C.~M., {Churchwell}, E., \& {Terzian}, Y. 1981, \aap, 96, 278

\bibitem[{{Walter}(1991)}]{1991PASP..103..830W}
{Walter}, D.~K. 1991, \pasp, 103, 830

\bibitem[{{Wang} {et~al.}(2017){Wang}, {Yu}, {Jiang}, {Zhao}, {Sun}, {Li},
  {Zhong}, {Dong}, {Michael}, {Xia}, {Zuo}, {Gou}, {Guo}, {Lu}, {Liu}, {Fan},
  {Jiang}, \& {Qian}}]{2017AcASn..58...37W}
{Wang}, J.~Q., {Yu}, L.~F., {Jiang}, Y.~B., {et~al.} 2017, Acta Astronomica
  Sinica, 58, 37

\bibitem[{{Wilson} {et~al.}(1997){Wilson}, {Filges}, {Codella}, {Reich}, \&
  {Reich}}]{1997A&A...327.1177W}
{Wilson}, T.~L., {Filges}, L., {Codella}, C., {Reich}, W., \& {Reich}, P. 1997,
  \aap, 327, 1177

\bibitem[{{Zhang} {et~al.}(2021){Zhang}, {Xu}, {Li}, {Hou}, {Yu}, \&
  {Jiang}}]{2021RAA....21..209Z}
{Zhang}, C.-P., {Xu}, J.-L., {Li}, G.-X., {et~al.} 2021, Research in Astronomy
  and Astrophysics, 21, 209

\bibitem[{{Zhu} {et~al.}(2022){Zhu}, {Wang}, {Zhu}, \&
  {Zhang}}]{2022A&A...665A..94Z}
{Zhu}, F.~Y., {Wang}, J.~Z., {Zhu}, Q.~F., \& {Zhang}, J.~S. 2022, \aap, 665,
  A94

\end{thebibliography}
\bibliographystyle{aa}


\end{CJK*}
\end{document}